\documentclass[fleqn,usenatbib]{mnras}

\usepackage{newtxtext,newtxmath}

\usepackage[T1]{fontenc}

\DeclareRobustCommand{\VAN}[3]{#2}
\let\VANthebibliography\thebibliography
\def\thebibliography{\DeclareRobustCommand{\VAN}[3]{##3}\VANthebibliography}

\usepackage{graphicx}	
\usepackage{amsmath}	
\usepackage{supertabular}
\defcitealias{Weltevrede2006}{W06}
\defcitealias{Malofeev2018}{M18}

\title{Search of RRATs on declinations from $+42^{\circ}$ to $+55^{\circ}$ with a neural network}

\author[I. V. Eldarov et al.]{
	I. V. Eldarov,$^{1}$
	S. A. Tyul'bashev,$^{2}$\thanks{E-mail: serg@prao.ru (SAT)}
	M. A. Kitaeva,$^{2}$
	G.E. Tyul'basheva $^{3}$
	\\
	$^{1}$ Russian Biotechnological University, ROSBIOTECH, Moscow region, Pushchino, Russia\\
	$^{2}$ Lebedev Physical Institute of RAS, Pushchino Radio Astronomy Observatory, Moscow region, Pushchino, Russia\\
	$^{3}$ Institute of Mathematical Problem of Biology, the branch of Keldysh Institute of Applied Mathematics of RAS, Moscow region, Pushchino, Russia\\
}

\date{ }

\pubyear{ }

\begin{document}
	\maketitle
	
	\begin{abstract}
In an area of 3,300 square degrees, a search for pulsed dispersed signals using a neural network has been carried out. During the six-month observation period, pulses were detected from fifteen known pulsars as well as three new rotating radio transients (RRATs). The main characteristics of these new sources were provided. The dispersion measures of the transients and the half-widths of the pulses ranged from 7.2 to 59.9 pc/cm$^3$ and from 20 to 300 ms, respectively. A RRAT search scheme has been developed that allows for the detection of pulses with a signal-to-noise ratio (S/N) below the threshold required for reliable detection.
	\end{abstract}

	\section{Introduction}
	
In 2003, a method was developed to search for short-duration signals that pass through the interstellar medium (\citeauthor{Cordes2003}, \citeyear{Cordes2003}). As an example of such signals, giant pulses from pulsars were considered. The idea behind the search was that signals passing through the medium are dispersed, meaning they arrive at the observer first at a high frequency and then at a low frequency, making them distinguishable from noise. By compensating for the time shifts of the signal at different frequencies and then combining the frequency channels, the original signal can be reconstructed.

In 2006, rotating radio transients (RRATs) were discovered (\citeauthor{McLaughlin2006}, \citeyear{McLaughlin2006}), and their signals were also found to be dispersed. It was immediately clear that RRATs were pulsars with unusual properties. The time intervals between successive pulses of these transients can vary from tens of seconds to hundreds of hours (\citeauthor{McLaughlin2006}, \citeyear{McLaughlin2006}; \citeauthor{Tyulbashev2024}, \citeyear{Tyulbashev2024}), but the standard periodic signal characteristic of canonical pulsars is often not detected between pulses even at very high fluctuation sensitivity (\citeauthor{Zhou2023}, \citeyear{Zhou2023}). However, RRAT periods that are multiples of the intervals between detected pulses are typical for slow (second) pulsars. We also note that according to the ATNF Pulsar Catalog (\citeauthor{Manchester2005}, \citeyear{Manchester2005})\footnote{https://www.atnf.csiro.au/research/pulsar/psrcat/}, periods are not determined for approximately 1/3 of the RRATs included in the catalog.

Initially, RRAT searches were performed using archived data from pulsar surveys (see, for example, \citeauthor{Keane2010}, (\citeyear{Keane2010}; \citeauthor{Burke-Spolaor2011}, (\citeyear{Burke-Spolaor2011}); \citeauthor{Karako-Argaman2015}, (\citeyear{Karako-Argaman2015}). Subsequently, standard programs were developed for searching for dispersed pulses, and pulsar searches began to include periodic and pulsed emission searches (\citeauthor{Keane2018}, \citeyear{Keane2018}; \citeauthor{Han2025}, \citeyear{Han2025}; \citeauthor{Turner2025}, \citeyear{Turner2025}). The typical duration of a pulsar search session is 5-10 minutes, so finding RRAT pulses is a side task. Given the possible time difference between consecutive RRAT pulses, which can range from tens of minutes to hundreds of hours, it is clear that many RRATs will be missed when using conventional pulsar search programs, as the time of arrival of the pulse is unpredictable.

Different methods are used to search for pulsar pulses. The main (standard) method of searching for RRAT, proposed in \citeauthor{Cordes2003}, (\citeyear{Cordes2003}), consists of various variations of iterating through dispersion measures ($DM$) by means of time shifts in frequency channels, followed by their addition and estimation of the observed signal-to-noise ratio (S/N) in the pulse on these $DM$. As a rule, if the signal found has S/N > 6-7, it is considered that the pulse has been detected. Almost all known RRATs were found in this way.

A fundamentally different approach was proposed in \citeauthor{Swinbank2015}, (\citeyear{Swinbank2015}). By constructing a series of VLBI maps of a region of the sky at short time intervals (typically a few seconds or tens of seconds), a transient can be detected as a blinking point on the maps. In fact, a pulse is an object whose intensity differs from the background radiation at a fixed point in the sky. Therefore, the image of a sky region with a pulse will differ from the image without a pulse. This is how a transient lasting several minutes was detected at a frequency of 60 MHz (\citeauthor{Stewart2016}, \citeyear{Stewart2016}), as well as several long-period transients (\citeauthor{Hurley-Walker2022}, \citeyear{Hurley-Walker2022}; \citeauthor{Caleb2024}, \citeyear{Caleb2024}).

Another way to search for pulsed radiation is to use neural networks (\citeauthor{Tyulbashev2022}, \citeyear{Tyulbashev2022}). If you create a dynamic spectrum, which is an image where the vertical axis represents frequency and the horizontal axis represents time, and the intensity of the signal in each pixel is represented by color, the pulsar's pulse will appear as a slanted line on the spectrum. You can train a neural network to search for these lines in images that represent dynamic spectra.

Regardless of the search method used, the final verification of the detected signals is performed visually, as the variety of interference makes it impossible to rely solely on formal signal detection.

There is no separate catalog for all detected RRATs. We estimate that approximately 270-300 transients are currently open. In February 2025, the ATNF catalog lists 211 RRATs. The vast majority of RRATs (228 sources) have been detected using standard search methods and published in 10 papers (\citeauthor{McLaughlin2006}, \citeyear{McLaughlin2006}; \citeauthor{Keane2010}, \citeyear{Keane2010}; \citeauthor{Burke-Spolaor2011}, \citeyear{Burke-Spolaor2011}; \citeauthor{Karako-Argaman2015}, \citeyear{Karako-Argaman2015}; \citeauthor{Tyulbashev2018}, \citeyear{Tyulbashev2018}; \citeauthor{Zhou2023}, \citeyear{Zhou2023}; \citeauthor{Tyulbashev2023}, \citeyear{Tyulbashev2023}; \citeauthor{Dong2023}, \citeyear{Dong2023}; \citeauthor{Tyulbashev2024}, \citeyear{Tyulbashev2024}; \citeauthor{Han2025}, \citeyear{Han2025}).

RRAT searches show that there are factors that greatly affect the ability to detect new transients. The main factors are the sensitivity of the radio telescope, the coverage area, and the duration of signal accumulation for a selected direction in the sky. In addition to new and previously discovered transients, searches also detect pulses from regular pulsars. Some of these pulses have a S/N ratio of < 6, but there is no doubt that they belong to pulsars because the time intervals between these weak pulses are multiples of the period.

Therefore, when searching for transients, it makes sense to set the RRAT candidate detection threshold at lower S/N values than is commonly accepted. However, our search experience shows that as the S/N level decreases, the number of transient candidates and the amount of interference increase dramatically, so pulses with S/N < 6 can only be used as additional confirmation of transient detection or to estimate the rotation period.

If there is a reliable way to filter out interference, it is possible to try to detect weaker RRATs. On a dynamic spectrum, they should appear as a faint dispersion delay line, unlike the usual RRATs observed at S/N > 7. To increase reliability, it is possible to use multiple detection of pulses in the same $DM$. These pulses should fall within the same area on the sky, with an angular size that matches the size of the antenna's beam pattern.

In two works on the search for transients using the standard search technique and the subsequent use of a neural network to filter out interference, the network was shown to work reliably with small losses of real transient pulses (\citeauthor{Tyulbashev2022}, \citeyear{Tyulbashev2022}; \citeauthor{Tyulbashev2024}, \citeyear{Tyulbashev2024}). In these works, the first stage of observation processing involved a standard search with an iteration of the $DM$ search and the fixation of the maximum value of the S/N.

In the PUMPS (Pushchino Multibeam Pulsar Search) survey data, approximately one million RRAT candidates are detected after processing one year of observations. Analysis shows that 75\% of the candidates are of various nature interferences (\citeauthor{Tyulbashev2022}, \citeyear{Tyulbashev2022}). The overwhelming part of the remaining candidates are pulses of known pulsars. For the found candidates the neural network acted as a filter that, in the second stage of work, selected reliable sources and significantly reduced the time for visual inspection of the remaining candidates in RRAT. In this work, we tested the neural network as a direct method for finding RRAT, including the search for weak transients with S/N < 6.

\section{Observations and processing}

The PUMPS survey (\citeauthor{Tyulbashev2022a}, \citeyear{Tyulbashev2022a}) is carried out on the Large Phased Array (LPA) radio telescope of the Lebedev Physics Institute (LPI). Two independent radio telescopes have been created based on an antenna field consisting of 16,384 wave dipoles covering an area of $200\times 400$ meters, and there is a technical possibility of creating two more radio telescopes.

The first radio telescope, LPA1, has a controllable beam pattern and a pulsar receiver that allows to change the number of frequency channels and the point sampling time (sampling frequency) within wide limits. Standard observations on LPA1 are carried out in a single beam.

The second radio telescope, LPA3, is equipped with a stationary system of 128 beams arranged in the meridian plane and covering declinations from $-9^{\circ}$ to $+55^{\circ}$. LPA3 is used as a monitoring telescope in the PUMPS project and in the "Space Weather" project (\citeauthor{Shishov2016}, \citeyear{Shishov2016}). The total reception bandwidth is 2.5 MHz. The LPA3 beam signal recording has fixed frequency channel widths (78 and 415 kHz), point sampling times (12.5 and 100 ms), and is recorded on three recorders.

A neural network was used for the search, which was trained using previously found pulses from pulsars and transients. A neural network (hereinafter referred to as a neural network/network) takes as input two-dimensional arrays of data with a size of 32 by 32 floating-point numbers, i.e., dynamic spectra of the same size, and returns probabilities. These probabilities indicate whether the input data matches or does not match the true signal. In other words, we obtain information in the form of a candidate for transients "found" / "not found". Thus, the neural network implements a mathematical task of binary classification of digital telescope data.

The neural network is implemented in the Python programming language, using the basic classes of the PyTorch framework\footnote{https://pytorch.org}, and consists of four neural layers connected by the ReLU (Rectified linear unit) activation function. The output layer has two neurons that determine the binary response to whether the signal belongs to the true class or not.

The data for training the neural network was prepared in two stages:

1) At the first stage, 4,500 pulses of pulsars and transients found in early searches (true-positive pulses) were selected. Then, a software augmentation procedure was performed, i.e., increasing the volume of data by creating copies with small modifications, where the original signals were stretched and compressed along the time and frequency axes, resulting in an increase of the positive class data volume for training to 45,000 elements.

The negative class data (false-positive pulses), i.e. dynamic spectra that do not contain candidate transients, were obtained by visual selection from the telescope data. A total of $\sim$205,000 false signals were selected. Thus, approximately 250,000 elements were used to train the neural network in the first stage.

2) The trained neural network of the first stage was used for a test search in data of one calendar month of observation. The search results were checked visually, which allowed to determine true-positive and false-positive signals. The volume of true-positive signals was $\sim$40 thousand, and false-positive $\sim$ 400 thousand elements. Thus, the new iteration allowed to carry out retraining of the network.

After the two stages, the secondary dataset was obtained by combining the primary dataset and the results of the first stage's neural network, followed by data augmentation of the true-positive and negative classes.

The augmentation procedure takes as input the maximum number of instances that need to be prepared for each $DM$ (signals with $DM$ ranging from 2 to 90 pc/cm$^3$ were generated), and the data was stretched and compressed based on the initial and target $DM$, as well as the specified maximum stretching/compressing threshold.

As a result of the augmentation, the volume of positive class data was $\sim$ 6 million instances, with a uniform distribution of signals and values of $DM=2-90$ pc/cm$^3$.

The negative class data was formed from the false-positive samples of the first stage neural network, and their augmentation consisted of sampling the telescope data with a small offset (forward and backward) from the coordinates of the original false-positive signal on the time axis, resulting in a training dataset of approximately 8 million samples. The second stage neural network was trained on approximately $\sim$6 million positive and $\sim$8 million negative class samples, for a total of 14 million samples.

To summarize the work done to train the neural network: stochastic gradient descent (SGD; \citeauthor{Robbins1951}, (\citeyear{Robbins1951})) was used for training; cross-entropy (\citeauthor{Shannon1948}, \citeyear{Shannon1948}; \citeauthor{Kiefer1952}, \citeyear{Kiefer1952}) was used as a loss function; the Adam (\citeauthor{Kingma2014}, \citeyear{Kingma2014}) optimizer was used to speed up the work; the number of training epochs was 80-100 (\citeauthor{Kullback1951}, \citeyear{Kullback1951}); the Learning Rate was set as 0.0001; the number of signals processed by the network in one call (Batch size) was 256,000; the network stopped training if there were 5 passes without changing the accuracy.

The trained neural network works under the control of a module that generates a classification task and groups positive signals that are found close enough to each other in the sky. The module that controls the neural network has a continuous search mode, where the search is conducted sequentially from the beginning to the end of the hour-long recording, and a targeted search mode that allows you to limit the right ascensions and declinations where you want to search. The second mode can be used to search for pulses from known pulsars and transients.

After training the neural network, a search was conducted in a six-month recording at declinations ranging from $+42^{\circ}$ to $+55^{\circ}$. Previously, the search for pulsed dispersed signals in this area was conducted to study the nature of pulsed interference (\citeauthor{Samodurov2022}, \citeyear{Samodurov2022}; \citeauthor{Samodurov2023}, \citeyear{Samodurov2023}). In these studies, it was shown that the recordings contain interference with characteristics of transients, pulses from 10 known pulsars, and pulses from six new RRATs.

\section{Detection of weak pulses of known pulsars}

\begin{figure}
	\centering
	\includegraphics[width=0.95\columnwidth]{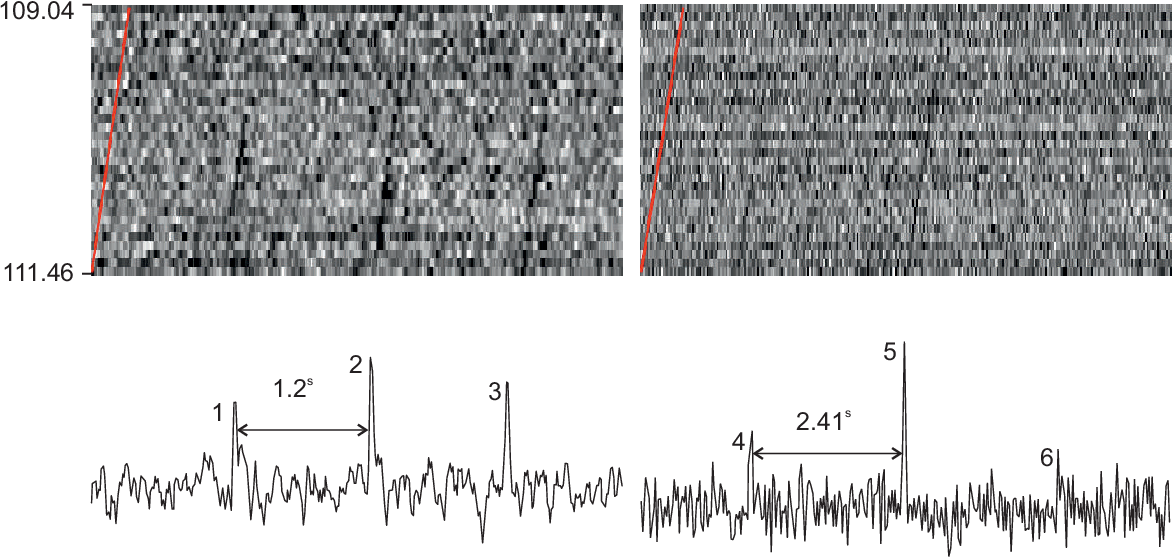}
	\caption{The figure shows the dynamic spectra and profiles of six pulses found by the neural network in the data from October 14, 2015. The vertical scales of the dynamic spectra and pulse profiles display the observation frequencies and intensities. The horizontal scale represents time. The values of the S/N found by the network show that the network's sensitivity in searching for the given patterns is high. Despite the fact that the neural network detected only a quarter of the visually identifiable pulses (mostly S/N<6), the network recognized some of the pulses with S/N < 5. These S/N values are lower than in previous search attempts (\citeauthor{Tyulbashev2018}, \citeyear{Tyulbashev2018}; \citeauthor{Tyulbashev2023}, \citeyear{Tyulbashev2023}), where the search was limited to pulses with S/N > 6-7.}
	\label{fig1}
\end{figure}

The neural network was configured in such a way that could "see"\,  faintly pronounced lines of dispersion delays. A vivid example of the network's operation when it is tested on known pulsars is shown in Fig. \ref{fig1}. PSR J0302+2252 is one of the first slow pulsars discovered in Pushchino (\citeauthor{Tyulbashev2016}, \citeyear{Tyulbashev2016}). In 2024, its parameters were refined using archived data from Arecibo (\citeauthor{Deneva2024}, \citeyear{Deneva2024}), and it was shown that its $P = 1.207165$~s, $DM = 18.99$~pc/cm$^3$, and the half-width of the pulses at 0.5 and 0.1 amplitude levels were estimated in Arecibo as $W_{0.5} = 30.2$~ms, $W_{0.1} = 76.6$~ms.

In the data from October 14, 2015, the network found 6 pulses. The pulsar's period is known, and a visual search of the original data revealed approximately 25 pulses after compensating for $DM$. The dynamic spectra (Fig. \ref{fig1}, top panel) show examples of pulses identified and non-identified by the neural network. On the left and right of the panels with dynamic spectra, three dispersion delay lines can barely be seen. The red color on the spectra shows $DM = 20$ pc/cm$^3$, where the maximum of the collected profiles was observed. The bottom panel of Fig. \ref{fig1} shows numbered pulses with a spacing of 1.2 and 2.41 seconds, which is a multiple of the pulsar's period. These pulses (from left to right) have a S/N of 4.5; 7.5; 6; 4.5; 9; 3. The neural network "saw"\, pulses 2, 3, 4, 5, but "did not see"\, pulses 1 and 6.

\begin{figure}
	\centering
	\includegraphics[width=1.0\columnwidth]{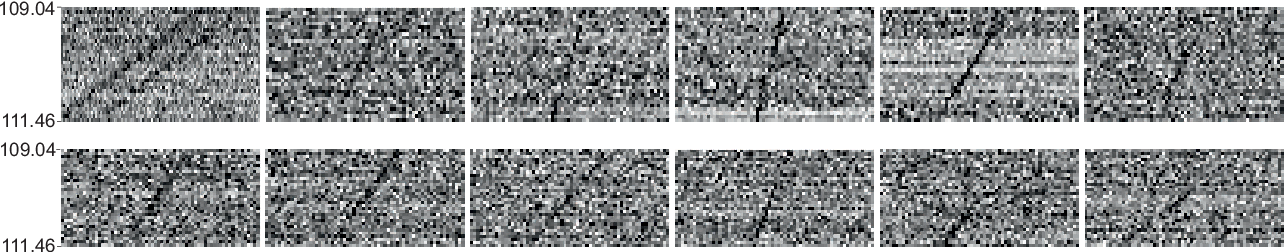}
	\caption{On the dynamic spectra of the upper panel, the RRAT candidates' pulses are shown, and on the lower panel, the dynamic spectra of the weak pulses of known pulsars are shown (from left to right J0014+4746, J0053+47, J0454+4529, J1115+5030, J1509+5531, B1753+52).}
	\label{fig2}
\end{figure}

Fig. \ref{fig2} shows the dynamic spectra of weak pulsar pulses found by the neural network and similar pulses from unidentified sources. In total, the network found pulses from 15 known pulsars. The main difference between the dynamic spectra in the top and bottom panels is that known pulsars have many pulses, while new transient candidates typically have only one pulse.

As can be seen from Fig. \ref{fig2}, the dynamic spectra of pulsar pulses and transient candidates do not differ visually. However, when checking weak single pulses of pulsars, we can rely on their known periods and dispersion measures, providing additional criteria for selecting pulses. For RRAT candidates, we do not have additional selection criteria \textit{a priori}, and it is necessary to introduce other physical criteria to determine whether a candidate is a new RRAT.

Next, we do not consider the pulses of known pulsars, as they are not the subject of this work. However, we provide on the website\footnote{https://bsa-analytics.prao.ru/en/transients/pulsars/}, which contains the results of the search for pulsars and transients by the PUMPS project, examples of dynamic spectra and pulse profiles found by the pulsar network (J0014+4746; B0053+47; B0329+54; J0454+4529; J1115+5030; J1509+5531; J1628+4406; B1753+52; J1923+4243; J1955+5059; J2022+5154; J2108+4516; J2113+4644; J2219+4754; J2313+4253).

\section{Search of new RRAT}

As shown in the previous paragraph (see Fig. \ref{fig1}), when searching for transient pulses, the neural network detects weak (S/N < 6) pulses from known pulsars. A natural question is how much we can trust similar-looking single pulses that do not belong to known pulsars or RRATs. To answer this question, we need to solve two problems. First, we need to consider the number of false detections of pulse signals at different S/N levels, assuming that we know the number of time intervals used for pulse search. Secondly, if several weak signals are detected in the LPA's directional pattern, can we say that we have discovered a new RRAT?

Assuming that the noise recorded on the LPA radio telescope has a normal distribution, it is possible to estimate the number of false detections of pulse signals at different S/N. By denoting the standard deviation of the noise as $\sigma_n$, it is possible to define the signal-to-noise ratio as $S/N = A/\sigma_n$ ($A$ is the amplitude of the signal). In this definition, the amplitude is expressed in terms of the noise signal.

To estimate the number of false detections, it is necessary to know the number of segments on which the pulse signals are checked. We have processed six-month observations at declinations from +42$^{\circ}$ to +55$^{\circ}$, which are covered by 32 LPA3 beams along the meridian. The search for pulses is carried out on segments of one second duration, with $DM$ ranging from 0 to 100 pc/cm$^3$. Thus, $24 \times 3600 \times 100 \times 32 = 2.76 \times 10^8$ segments are checked per day, and $N = 180 \times 2.76 \times 10^{8} = 5 \times 10^{10}$ segments are checked per six months. The probability ($p$) of detecting a false signal at a level above 5, 6, and 7$\sigma_n$ is equal to $p = 5.7 \times 10^{-7}; 2.0 \times 10^{-9}; 2.56 \times 10^{-12}$. Thus, in six months of observations, for signals with $A$ = 5, 6, 7$\sigma_n$ (S/N = 5, 6, 7), it is possible to detect $p \times N = 2.8 \times 10^4; 100; 0.13$ false sources.

The probability of 0.13 (or 13\%) to detect a false pulse at the level of S/N = 7 seems small, however, at the moment about 300 RRATs are already known (see Introduction). If all these RRATs would be detected on LPA at the level of S/N = 7, among them there would be a high probability of false detections. Therefore, when interpreting the search results, it is necessary to introduce such a level of S/N, at which a single detection of a transient would speak with a high probability of the reality of the found transient. In this work, we will adhere to the criterion that a source is true if the probability of a false detection is less than 0.001.

It is obvious that the possible tens of thousands of false detections of transients are an absolutely unacceptable situation when searching for RRAT. However, the survey area is 3300 sq. deg. or 7000 independent directions in the sky, taking into account the size of the LPA's radiation pattern. False sources fall into this area randomly, and the probability of two, three, and so on sources falling into one direction (one LPA beam) may be low. It is possible to calculate the probabilities of such random hits, assuming that a Poisson distribution can be used for random (false) pulses. In this case, if the probability of hitting one direction of several pulse signals is small, we can talk about the detection of a new transient.

It should also be noted that RRAT pulses are detected at different $DM$. This means that in addition to multiple false sources being detected in the same LPA beam, they must also have the same dispersion measure. The typical accuracy of determining $DM= \pm 2.5$ pc/cm$^3$ is up to 100 pc/cm$^3$ (see, for example, \citeauthor{Tyulbashev2018}, (\citeyear{Tyulbashev2018})), so the total number of pulses for which we will consider the probability of accidentally detecting several of them in the same direction should be reduced by a factor of 100/5 = 20.

As we can see, the main quantity that determines the reality of a detected transient is its S/N. The neural network "guesses"\, the pulse, but does not tell us about its S/N. To obtain an estimate of the S/N, we perform a search over $DM$, add the frequency channels, subtract the baseline, estimate $\sigma_n$, and compare $\sigma_n$ with the signal amplitude.

In a regular search, additional averaging of the time array with different steps is performed for a candidate transient. The averaging allows to integrate the pulse energy and increase the observed S/N. The median half-width of the pulses detected on LPA transients, according to the search results table\footnote{https://bsa-analytics.prao.ru/transients/rrat/}, is 23 ms. By averaging the pulse profile at different time steps, it is possible to increase the S/N ratio. The pulse shape of most detected transients looks like a triangle. Given that the sampling time is 12.5 ms, a typical increase in the S/N ratio when averaging over 4 points is approximately 30-40\%. If the pulse is described by a Gaussian shape, the S/N ratio gain from averaging will be higher. To maximize the S/N ratio of the RRAT candidates, we performed averaging using a moving average filter with a step size that covers a wide range of pulse widths.

In this work, we obtained several levels of the S/N, depending on which we considered a reliable RRAT discovery when the pulses were detected once, twice, and three times. It turned out that for the S/N $\ge 7.7$, a single detection is sufficient to consider the discovery reliable. If 5.3 $\le$ S/N < 7.7, a double detection is necessary for a reliable discovery. If 5.0 $\le$ S/N < 5.3, a triple detection is required.

The obtained estimates show that the boundaries of the S/N for a three-fold random detection are narrow, and their use in practice does not make sense. We also note that the given estimates concern the declination values we have chosen and the half-yearly daily round-the-clock observations on the LPA. When changing the number of processed days and the size of the studied area, the boundaries of the S/N levels for reliable detection of sources must be recalculated.

After removing the pulses of known pulsars and the pulses that did not pass the visual inspection, there were only 14 candidates for new RRATs. No candidate had two or more pulses in the same LPA beam. Using a moving average filter, we integrated the energy of the pulses and estimated the S/N of the RRAT candidates. In Table 1, we present only those RRATs with a false detection rate of less than 0.001 (S/N $\ge$ 7.7). Columns 1-3 show the name of the transient and its RA and DEC coordinates as of 2000. The coordinates are accurate to $\pm1.5^m$ and $\pm15'$. Columns 4-7 show the estimates of $DM$, peak flux density $S_p$, half-width of the pulse $W_{0.5}$, and the S/N of the pulse before and after moving average averaging. For J0840+42, the pulse was detected in two adjacent beams. The peak flux density is estimated after compensation for inter-beam hits, and the remaining characteristics are averaged. Dynamic spectra and pulse profiles are available on the website\footnote{https://bsa-analytics.prao.ru/en/transients/rrat/}.

\begin{table*}[]
	\caption{
    Characteristics and coordinates of the detected RRAT
    }
	\begin{tabular}{|c|c|c|c|c|c|c|}
		\hline
name       &   $\alpha_{2000}$  &   $\delta_{2000}$ &  $DM$ (pc cm$^{-3}$) &   $S_p$ (Jy) &   $W_{0,5}$ (ms)  & S/N\\
\hline
J0101+46  &   $01^h01^m40^s$   &  $46^{\circ}42'$  &  59.9$\pm$1.0     &   2.6        &   300   &    7.0 (10)\\
J0841+46  &   $08^h41^m37^s$   &  $46^{\circ}15'$  &  7.2$\pm$0.5        &   5.0        &    20   &     19 (19)\\
J2053+50  &   $20^h53^m18^s$   &  $50^{\circ}50'$  &  15.6$\pm$0.5       &   6.0        &    25   &    7.3 (7.9)\\
\hline
\end{tabular}
\label{tab:1}
\end{table*}

As can be seen from Table 1, the RRATs found have a S/N > 7.7. The remaining sources have a S/N < 7.7, and therefore these RRAT candidates were discarded.

\section{Discussion and Conclusion}

As shown in the previous section, the search should detect thousands of false sources with S/N $\sim$ 5-6. This is exactly what happened. The network found thousands of weak sources that were discarded during visual inspection. It turned out that the false transient, formally detected by the neural network, "scanning" dynamic spectra, is very different from the dispersion pulses of real pulsars and transients. This allows us to filter out false transients and reduce the number of candidates by hundreds after excluding known pulsars.

Let's consider what a typical strong transient looks like in dynamic spectra. Most often, it appears as a straight line that intersects all or almost all frequency channels at an angle. In reality, the dispersion delay line (the amount of time offset between the signal and the observation frequency) has a parabolic shape, but due to the narrow bandwidth of the LPA receiver, this parabolic section is indistinguishable from a straight line. Dynamic spectra often show bands where the intensity in pixels along the dispersion delay line drops significantly (see, for example, \citeauthor{Tyulbashev2018}, (\citeyear{Tyulbashev2018}). These bands are likely caused by the fact that pulsar pulses may have linear polarization, resulting in Faraday rotation of the polarization plane.

For weak pulses, the picture becomes blurred. In their dynamic spectra, there may be areas where the signal's dispersion delay line is visible, and areas where the intensity drops to the point of signal disappearance. In some areas, the intensity in the pixels does not differ from the background intensity. For both strong and weak transients, the dispersion delay line has approximately the same width when viewed along the line, and it consists of segments of equal length rather than spots.

Of course, the "spotted" pulse structure in the dynamic spectrum cannot be completely excluded. Such structures are often observed in the dynamic spectra of pulsars if there are diffraction interstellar scintillations. In this case, a characteristic "spotted" pattern appears in the dynamic spectra, where the pulse energy varies on time and frequency scales. However, with the 78 kHz frequency channel width used in the observations, the appearance of such structures is unlikely. When visually inspected, pulses consisting of spots can be excluded. At the same time, by adding energy along the "line" the dispersion delay (by adding the energy of the spots), an pulse can be obtained that is externally indistinguishable from a regular transient pulse.   

The normal (gaussian) distribution tells us that if we check time segments from chopped-up noise and look for signals with a S/N greater than a specified value, then with an infinite number of segments, we can find a signal with an arbitrarily large S/N. In our case, the normal distribution "does not know" anything about how the segment is formed, where the pulses are searched.

\begin{figure}
	\centering
	\includegraphics[width=0.7\columnwidth]{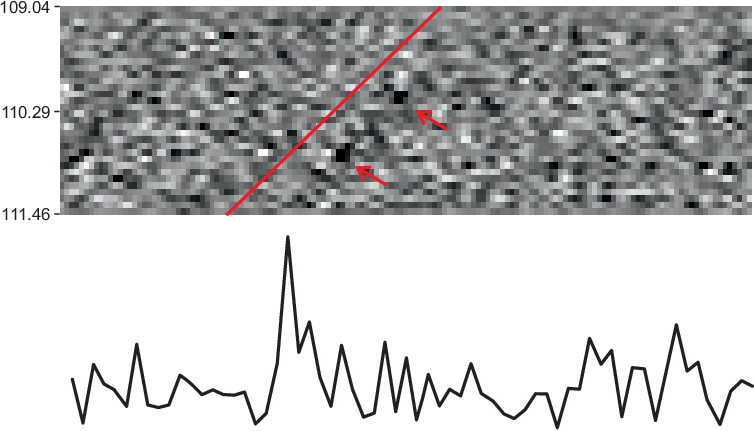}
	\caption{In the dynamic spectrum (top panel), the arrows indicate the pixels of increased brightness (spots) that lie "parallel" to the dispersion delay line (red line) corresponding to $DM = 19$ pc/cm$^3$. The bottom panel shows the profile of the "pulse" stacked on this dispersion measure.}
	\label{fig3}
\end{figure}

A false pulse is formed by separate independent bright points or spots (pixels) in the frequency channels. The number of these points may vary. In the dynamic spectrum, they lie approximately along the dispersion delay line. The shape of the pulse profile depends on the distribution of energy in the spots. An example of such a noise signal is shown in Fig. \ref{fig3}. In fact, almost all the energy came from two spots and formed the profile after compensating for the dispersion delay.

When the dynamic spectra of the found candidates were initially viewed, such "pulses" were immediately discarded, as no real dispersion delay line was visible in the spectrum, and the pseudoline was a visual illusion formed by multiple spots that disappeared when the image was enlarged. Faraday rotation also failed to explain the apparent energy distribution along the line, as no regular structures of high-brightness pixels were observed.

The mass rejection of transient candidates has led to the fact that we are only writing about the verification of 14 pulses. The total number of patterns that the neural network is trained on is small. This is due to the fact that so far, we have detected only about 100 RRATs and about 125 pulsars in the LPA3 monitoring data\footnote{https://bsa-analytics.prao.ru/transients/}. This means that the training sample is very small. It is possible that introducing artificial pulses with different $DM$, intensity, and half-width into the original data will allow the network to be retrained and improve the rejection of false objects, thereby reducing manual labor.

Let's summarize the method of searching for pulse dispersion signals, which includes the search for weak pulses. It should consist of the following steps:

- search for candidates in RRAT by a neural network, and save their dynamic spectra;

- initial visual inspection, and filtering out obviously false candidates;

- evaluation of the dispersion measure, and evaluation of the pulse's S/N;

- depending on the obtained S/N, we remember the candidate in RRAT for the case of single detection (S/N $\ge 7.7$) and double detection on the same $DM (5.3 \le$ S/N $< 7.7$). The S/N boundaries will shift depending on the number of segments checked;

 -final visual inspection.

In conclusion, we note the main results. The neural network allowed us to search for weak (S/N < 6) signals that have a dispersion delay in the frequency channels. The total number of candidates checked was reduced by dozens of times compared to standard search methods. The majority of new candidates in RRAT were objects that have a statistical nature. It is necessary to develop a method for their automatic removal before visual inspection. Three new rotating radio transients have been discovered.

\section*{Acknowledgements}

The study was carried out at the expense of the Russian Science Foundation grant 22- 12- 00236-$\Pi$. (https:// rscf.ru/ project/ 22- 12- 00236-$\Pi$/). The authors are grateful to the antenna group that maintains the LPA radio telescope for their constant assistance.


\begin{thebibliography}{99}
\bibitem[Cordes \& McLaughlin(2003)]{Cordes2003} Cordes, J.~M. \& McLaughlin, M.~A., ApJ, 596, {\bf 1142}, (2003).
\bibitem[McLaughlin et al.(2006)]{McLaughlin2006} McLaughlin, M.~A., Lyne, A.~G., Lorimer, D.~R., et al., Nature, {\bf 439}, 817, (2006). 
\bibitem[Tyul'bashev et al.(2024)]{Tyulbashev2024} Tyul'bashev, S.~A., Kitaeva, M.~A., Pervoukhin, D.~V., et al., A\&A, {\bf 689}, A1, (2024). 
\bibitem[Zhou et al.(2023)]{Zhou2023} Zhou, D.~J., Han, J.~L., Xu, J., et al., Research in Astronomy and Astrophysics, {\bf 23}, 104001, (2023). 
\bibitem[Manchester et al.(2005)]{Manchester2005} Manchester, R.~N., Hobbs, G.~B., Teoh, A., et al., AJ, {\bf 129}, 1993, (2005). 
\bibitem[Keane et al.(2010)]{Keane2010} Keane, E.~F., Ludovici, D.~A., Eatough, R.~P., et al., MNRAS, {\bf 401}, 1057, (2010). 
\bibitem[Burke-Spolaor et al.(2011)]{Burke-Spolaor2011} Burke-Spolaor, S., Bailes, M., Johnston, S., et al., MNRAS, {\bf 416}, 2465, (2011). 
\bibitem[Karako-Argaman et al.(2015)]{Karako-Argaman2015} Karako-Argaman, C., Kaspi, V.~M., Lynch, R.~S., et al., ApJ, {\bf 809}, 67, (2015). 
\bibitem[Han et al.(2025)]{Han2025} Han, J.~L., Zhou, D.~J., Wang, C., et al., Research in Astronomy and Astrophysics, {\bf 25}, 014001, (2025). 
\bibitem[Keane et al.(2018)]{Keane2018} Keane, E.~F., Barr, E.~D., Jameson, A., et al., MNRAS, {\bf 473}, 116, (2018). 
\bibitem[Turner et al.(2025)]{Turner2025} Turner, J.~D., Stappers, B.~W., Tian, J., et al., MNRAS, {\bf 537}, 1070, (2025). 
\bibitem[Swinbank et al.(2015)]{Swinbank2015} Swinbank, J.~D., Staley, T.~D., Molenaar, G.~J., et al., Astronomy and Computing, {\bf 11}, 25, (2015). 
\bibitem[Stewart et al.(2016)]{Stewart2016} Stewart, A.~J., Fender, R.~P., Broderick, J.~W., et al., MNRAS, {\bf 456}, 2321, (2016). 
\bibitem[Hurley-Walker et al.(2022)]{Hurley-Walker2022} Hurley-Walker, N., Zhang, X., Bahramian, A., et al., Nature, {\bf 601}, 526, (2022). 
\bibitem[Caleb et al.(2024)]{Caleb2024} Caleb, M., Lenc, E., Kaplan, D.~L., et al., Nature Astronomy, {\bf 8}, 1159, (2024). 
\bibitem[Tyul'bashev et al.(2022)]{Tyulbashev2022} Tyul'bashev, S.~A., Pervukhin, D.~V., Kitaeva, M.~A., et al., A\&A, {\bf 664}, A37, (2022). 
\bibitem[Tyul'bashev et al.(2018)]{Tyulbashev2018} Tyul'bashev, S.~A., Tyul'bashev, V.~S., \& Malofeev, V.~M., A\&A, {\bf 618}, A70, (2018). 
\bibitem[Tyul'bashev et al.(2023)]{Tyulbashev2023} Tyul'bashev, S.~A., Kitaeva, M.~A., Brylyakova, E.~A., et al., Astronomy Letters, {\bf 49}, 533, (2023). 
\bibitem[Dong et al.(2023)]{Dong2023} Dong, F.~A., Crowter, K., Meyers, B.~W., et al., MNRAS, {\bf 524}, 5132, (2023). 
\bibitem[Tyul'bashev et al.(2022a)]{Tyulbashev2022a}Tyul'bashev, S.~A., Kitaeva, M.~A., \& Tyul'basheva, G.~E. 2022, MNRAS, {\bf 517}, 1112, (2022). 
\bibitem[Shishov et al.(2016)]{Shishov2016} Shishov, V.~I., Chashei, I.~V., Oreshko, V.~V., et al., Astronomy Reports, {\bf 60}, 1067, (2016). 
\bibitem[Robbins and Monro(1951)]{Robbins1951} Herbert Robbins and Sutton Monro, "A Stochastic Approximation Method", 
The Annals of Mathematical Statistics, {\bf 22}, No. 3., pp. 400-407, (1951). 
\bibitem[Kiefer and Wolfowitz(1952)]{Kiefer1952}J. Kiefer and J. Wolfowitz, "Stochastic Estimation of the Maximum of a Regression Function", Ann. Math. Statist. {\bf 23}, Number 3, 462-466,  (1952).
\bibitem[Shannon(1948)]{Shannon1948} C. E. Shannon, "A mathematical theory of communication", The Bell System Technical Journal {\bf 27 (3)}. 379-423,  (1948).
\bibitem[Kingma \& Ba(2014)]{Kingma2014} Kingma, D.~P. \& Ba, J., arXiv:1412.6980 (2014). 
\bibitem[Kullback, Leibler(1951)]{Kullback1951} S. Kullback, R. A. Leibler, "Kullback-Leibler divergence, On Information and Sufficiency", Ann. Math. Statist. {\bf 22(1)}, 79-86, (1951) 
\bibitem[Samodurov et al.(2023)]{Samodurov2023} Samodurov, V.~A., Tyul'bashev, S.~A., Toropov, M.~O., et al., Astronomy Reports, {\bf 67}, 590, (2023). 
\bibitem[Samodurov et al.(2022)]{Samodurov2022} Samodurov, V.~A., Tyul'bashev, S.~A., Toropov, M.~O., et al., Astronomy Reports, {\bf 66}, 341, (2022). 
\bibitem[Tyul'bashev et al.(2016)]{Tyulbashev2016} Tyul'bashev, S.~A., Tyul'bashev, V.~S., Oreshko, V.~V., et al., Astronomy Reports, {\bf 60}, 220, (2016). 
\bibitem[Deneva et al.(2024)]{Deneva2024} Deneva, J. S., McLaughlin, M., Olszanski, T. E. E., Lewis, E. F., Pang, D., Freire, P. C. C., Bagchi, M. \& Stovall, K., The AO327 Drift Survey Catalog and Data Release of Pulsar Detections. ApJS, {\bf 271(1)}, 23, (2024).

\end{thebibliography}
\end{document}